\documentclass[]{aa}
\usepackage{times}
\usepackage[]{graphics}
\usepackage[]{psfig}

\begin{document}

   \thesaurus{}
   \title{Exploring the gravitationally lensed system \object{HE
          1104$-$1805}:\\ VLT spectroscopy of the lens at z=0.729
          \thanks{Based on observations collected during ESO Program
          65.O-0566(A)}}

   \author{C. Lidman
          \inst{1}
	  \and
	  F. Courbin
	  \inst{2}
	  \and
          J.-P. Kneib
          \inst{3}
          \and
          G. Golse
	  \inst{3}
	  \and
          F. Castander
          \inst{3}
          \and
          G. Soucail
          \inst{3}
          }

   \offprints{C. Lidman}

   \institute{ European Sourthern Observatory, Casilla 19, Santiago,
              Chile\\ email: clidman@eso.org 
              \and 
              Universidad Cat\'olica de Chile, Departamento de
              Astronomia y Astrofisica, Casilla 306, Santiago 22,
              Chile\\ email: fcourbin@astro.puc.cl
              \and 
              Laboratoire d'Astrophysique, Observatoire
              Midi-Pyr\'en\'ees, UMR5572, 14 Avenue Edouard Belin,
              F-31000, Toulouse, France\\ email: kneib@obs-mip.fr, 
              ggolse@obs-mip.fr, castander@obs-mip.fr, soucail@obs-mip.fr }

   \date{Received, 2000; accepted, 2000}

   \titlerunning{The redshift of the lens in \object{HE~1104$-$1805}}
   \authorrunning{C. Lidman et al.}	
   \maketitle

   \begin{abstract}
	
   Using FORS2, mounted on Kueyen (UT2 of the VLT), we have obtained
the redshift of the lensing galaxy in the gravitational lens system
\object{HE~1104$-$1805}. We measure $z=0.729 \pm 0.001$, in good
agreement with previous estimates based on the time delay and the
position of the lens on the fundamental plane.  It also coincides with
one of the metal line absorption systems that are seen in high resolution
spectra of \object{HE~1104$-$1805}.

   \keywords{gravitational lensing $-$
             quasars; individual: \object{HE~1104$-$1805} $-$
             data processing
            }
   \end{abstract}

%
%

\section{Introduction}

\object{HE~1104$-$1805} was discovered as part of the Hamburg/ESO
Quasar Survey and was first identified as a gravitational lens
candidate by Wisotzki et al. (\cite{Wizot93}).  It
consists of two lensed images of a radio-quiet quasar at $z$=
2.319 that are separated by $\sim$3.2\arcsec.  The lensing galaxy was
discovered from ground based near-IR (Courbin, Lidman \& Magain,
\cite{Courbin98a}) and HST optical observations (Remy et
al. \cite{Rem1998}; hereafter R98) and is 1\arcsec\, from the brighter quasar image
(component A).

From a spectrophotometric monitoring program that lasted several
years, Wisotzki et al. (\cite{Wizot98}; hereafter W98) measured a time
delay of 0.73 years between the two components. To convert this into
an estimate of the Hubble Constant, one needs to determine the
geometry of the system (astrometry and lens and source redshifts) and
to model the mass distribution of the lens. Several mass models for
the lens have been published (W98; R98; Courbin et
al. \cite{Courbin2000a}; hereafter C2000; Lehar et al. \cite{Lehar00})
and precise astrometry from HST images is available. Since the source
redshift is known, the remaining unknown is the redshift of the lens.

In their discovery paper, Wisotzki et al. (\cite{Wizot93}) noted that
the continuum of component A was considerably harder than that of
component B, and that the broad emission lines of the two components
were identical, only scaled by a factor of 2.8.  This was still the
case during the spectrophotometric monitoring carried out by Wisotzki
et al. (\cite{Wizot98}) and in IR spectra covering the 1 to 2.5~$\mu$m
range (C2000). This has been interpreted as microlensing of the
continuum emitting region of the quasar in component A, which is twice
as close to the lensing galaxy as component B. Presumably, the broad
line region is too large to be affected. It may be possible to use
this information to gain insight into the geometries and sizes of the
continuum emitting and broad line regions; however, detailed
predictions require the lens redshift.

From high resolution spectroscopic observations (Smette et al.
\cite{Smette95}; Lopez et al. \cite{Lopez99a}), several metallic
absorption line systems have been detected: $z=0.517, 0.728, 1.280,
1.320, 1.662, 1.860, 2.220$ and $z=2.300$. The systems at $z=0.728$,
$z=1.320$ and $z=1.662$ contain lines that are mostly detected in
component A, the component that is closest to the lens. At one time or
another, each system has been individually proposed to be the lens.  

Despite its importance, the redshift of the lensing galaxy has proved
elusive.  Apart from the many unpublished attempts to measure it, the
redshift of the lens has been estimated by indirect means.  From IR
and optical photometry, C2000 gave a redshift in the range $z=0.8$ to
$z=1.2$ while from the time delay and a model of the lens, W98
estimated $z=0.79$. From the position of the lens on the fundamental
plane, Kochanek et al. (\cite{Koe20}) derived $z=0.77$. Although model
dependent, the two latter estimates prove to be very close to the
truth.

In this paper we describe the successful measurement of the lens
redshift, $z=0.729$, using the spectral deconvolution method described
by Courbin et al. (\cite{Courbin2000b}).  

%
%

\begin{figure*}[t]
\begin{center}
\resizebox{18.0cm}{!}{\includegraphics{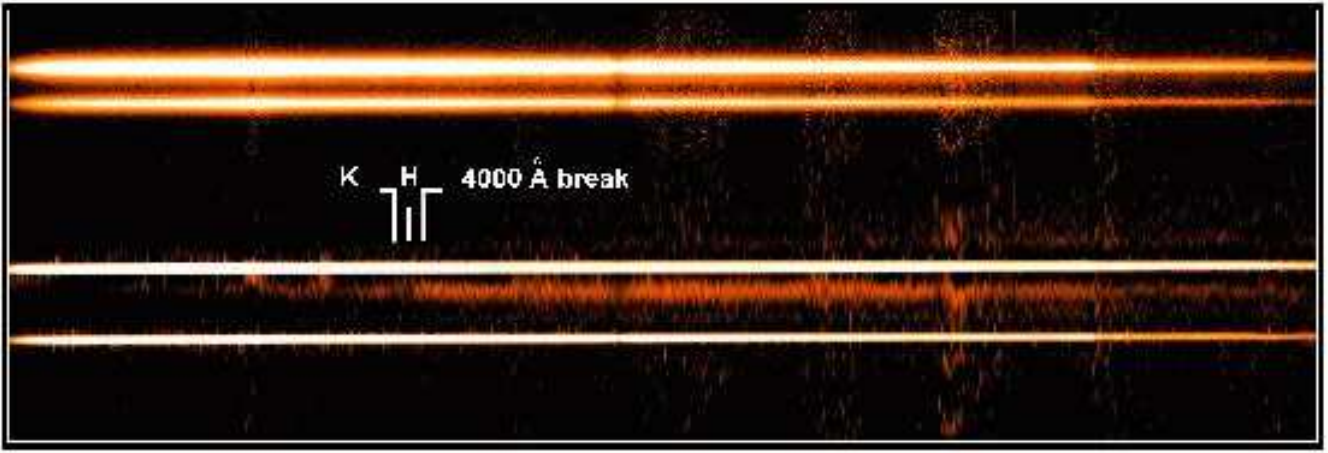}}
\caption[]{{\it Top:} the 2D sky-subtracted spectrum of
\object{HE~1104$-$1805}. This spectrum is the mean of the three 1080s
exposures. {\it Bottom:} deconvolved spectrum. The spectra of the quasar
images have a FWHM of 0.2\arcsec. The pixel size is half that of the
original data (hence, the height of this image is twice
that of the original), i.e., 0.05\arcsec. The main features of
the lensing galaxy are marked. The scale in the spatial direction is
given by the separation between the 2 quasar images, i.e.,
3.14\arcsec.}
\label{2d_spec}
\end{center}
\end{figure*}

\section{Observations-Reductions}

The observations were taken with FORS2 on Kueyen (VLT/UT2) at the Cerro
Paranal Observatory on April 1st, 2000, which also happened to be the
first scheduled observing night for this telescope, and consisted of
three 1080 second exposures with the G200I grism and the high
resolution collimator. This gives a scale of 0.1\arcsec\, per pixel in
the spatial direction and a scale of approximately 4\AA\, per pixel in
the spectral direction.  The observing conditions were clear and the
external seeing varied between 0.5\arcsec\, and 0.9\arcsec.

We used the movable slits in the FORS2 focal plane to select targets
and set the slit width.  One slit was placed on \object{HE~1104$-$1805} and
aligned along the two quasar components, three slits were placed on field
stars that were used to determine the PSF required for the
deconvolution (see below) and to correct for telluric absorption. Used
in this mode, the slits of FORS2 have a fixed length. This is less
flexible than punched or laser designed masks, but it was adequate for
our experiment.

The slit width was set to 1\arcsec\,, which is larger than the FWHM of
the images.  For the deconvolution to work well, it is important that
the targets are not too heavily occulted by the slit edges, since this
can lead to differential occultation between the PSF stars and the
quasar images and hence to a less accurate deconvolution.  This can
be minimized by carefully centroiding the stars in the slits and by
having slits that are significantly wider than the PSF. Additionally,
preparing the mask from an image that was taken with the same
instrument, as was done in this case, also minimises differential
alignment errors.

Instrumental signatures were removed in the standard manner using
IRAF. The bias was removed by subtracting a constant from each frame
and the pixel-to-pixel sensitivities were normalised with screen
flats.  The arc frames were used to correct for slit distortion and to
set the wavelength scale. All spectra were rebinned to the same linear
wavelength scale.  The position of the night sky lines and the
absorption features in component A were used to check the accuracy of
the wavelength calibration, which was found to be around $3\AA$.

\section{Spectral deconvolution: deblending the lens and source}

The lensing galaxy is four to five magnitudes fainter than component A
of \object{HE~1104$-$1805} and only one arc second away. Even with
excellent seeing, the spectrum of the lens is heavily contaminated
by that of the quasar and needs to be extracted with sophisticated
techniques.

The technique implemented by Courbin et al. (\cite{Courbin2000b})
spatially deconvolves spectra and decomposes them into point-sources
and extended components. It is therefore very well suited to the
present problem: extracting the spectrum of a faint extended lens
galaxy close to a bright quasar spectrum. As is the case with image
deconvolution, a reference PSF is necessary. Out of the three PSF
stars that were available, two were unfortunately too faint. In order
to build the PSF spectrum, we therefore used only one PSF star (which
was slightly fainter than the quasar itself) in combination with the
spectrum of the bright quasar image itself. Only half of the spatial
profile of the quasar was actually used, the one unaffected by light
contamination from the lensing galaxy. The deconvolved spectrum has a
pixel size of 0.05\arcsec\, while the undeconvolved data have a pixel
size of 0.1\arcsec. The deconvolved spectrum is displayed in
Fig.~\ref{2d_spec}.

As the PSF stars were observed at exactly the same time and under the
same conditions as the lensed quasar, they could also be used to
remove telluric features.  The spectra of the PSF stars were
combined and then divided directly into the spectra of the lens and
the two components of the quasar. Before division, the continuum of
the PSF stars was normalised with a high-order function and obvious
absorption features, such as ${\rm H}\alpha$ and the Calcium triplet
at 8600 \AA, were removed. As we do not know the spectral type of the
PSF stars, weaker features may be present.  Ideally, one should choose
hot, featureless stars to make such a correction.  Despite this, the
correction works rather well.

\begin{figure}[t]
\resizebox{9.0cm}{!}{\includegraphics{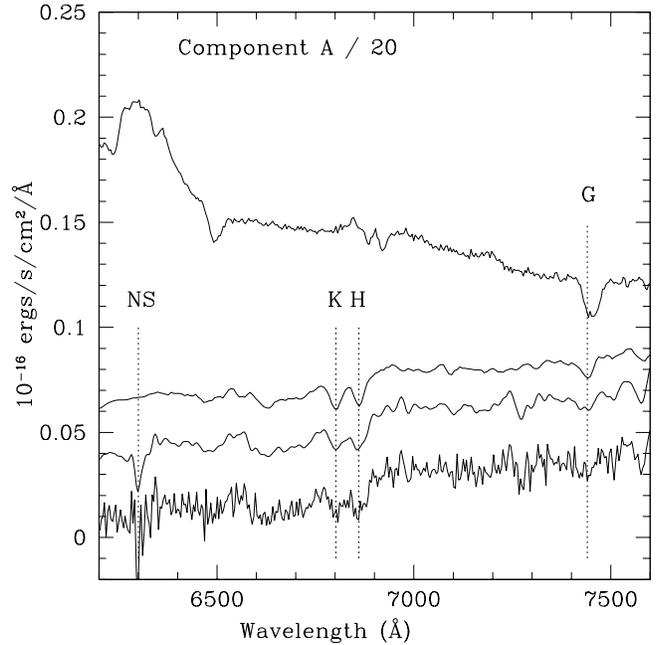}}
\caption[]{From bottom to top the curves are: an unsmoothed spectrum
of the lens in \object{HE~1104$-$1805}; a smoothed version with a
vertical offset of 0.03 units; the spectrum of the template galaxy
used in the cross correlation with a vertical offset of 0.05 units;
and the spectrum of component A of the quasar divided by 20. The
major absorption lines are identified with the dotted
lines. A night sky feature at 6300 $\AA$ is also marked.  The spectrum
of the lens shows no trace of contamination by the quasar emission
lines.}

\label{lens}
\end{figure}

The 1-D spectrum of the lens, extracted directly from the 2-D
deconvolved image (extended component only), is shown in
Fig.~\ref{lens}. An unsmoothed version and a smoothed version shifted
by 0.03 units on the vertical scale are shown in the lower half of the
figure.  Also shown, but reduced by a factor of 20, is the spectrum of
component A. The Calcium H and K absorption lines of the lensing
galaxy are clearly detected and the G-band is marginally detected.
The redshift was measured by cross correlating the lens spectrum with
the elliptical template from the Kinney-Calzetti spectral atlas
(Kinney et al. \cite{Kinney96}), which has been plotted (third curve
from the bottom) with a shift of 0.05 units in Fig.~\ref{lens}. For
this we used the IRAF task RVSAO v2.0 (Kurtz and Mink 1998).  The
redshift is $z=0.729 \pm 0.001$, where the error incorporates the
internal error reported by the cross-correlation routine and the error
in the wavelength calibration. An alternative measure of the accuracy
is given by the r-statistic of Tonry and Davis (\cite{Tonry79}),
which is the signal-to-noise ratio of the main peak in the cross
correlation. We find that $r=4.1$, which indicates that the redshift
is reliable (Kurtz \& Mink \cite{Kurtz98}).

The redshift is in good agreement with the redshift of the multiple
absorption line system at a redshift of $z=0.728$ (Smette et
al. \cite{Smette95}; Lopez, \cite{Lopez99b}). This absorption line
system is made up of many sub-components spread over a range of
redshifts ($\Delta z \approx 0.001$).  Although it is not possible to
say if the absorption line system and the lens are the same object,
the two systems are probably dynamically associated.

As the quality of the PSF was not optimal, the quasar spectra from the
deconvolved images are noisier than what one would calculate from the
photon noise only. Therefore, the extraction of the quasar spectra was
not done from the deconvolved spectrum but with narrow apertures from
the undeconvolved spectrum. Since the galaxy is much fainter than
either component of the quasar and since it is not spatially
coincident, the contamination of the quasar spectra by the lensing
galaxy is negligible.

\begin{figure}[t]
\resizebox{9.0cm}{!}{\includegraphics{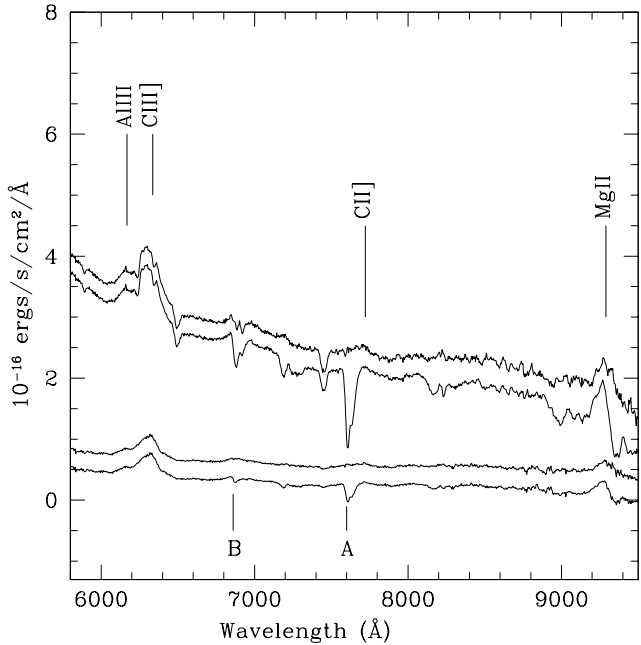}}
\caption[]{One dimensional spectra of components A and B of
\object{HE~1104$-$1805}. Both the spectra before and after the
correction for telluric features was applied are shown. For clarity,
the uncorrected spectra are shifted down by 0.3 units. The main quasar
emission lines and the atmospheric A and B bands are marked.}
\label{quasar}
\end{figure}

The spectra of the two components of the quasar are shown in Fig.~\ref{quasar}.
In this figure we plot the data before and after the telluric
features have been removed. The A and B atmospheric bands are removed
very well and the FeII absorption lines at $z=1.66$ become visible.
Additionally, the MgII emission line, which is almost obscured by the
strong telluric feature at 9300\AA, is recovered.

\begin{figure}[t]
\resizebox{9.0cm}{!}{\includegraphics{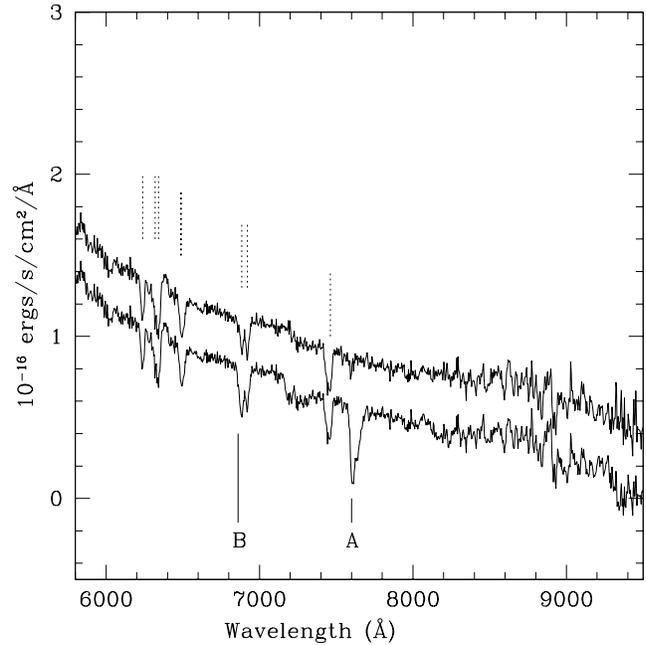}}
\caption[]{The difference between the spectra of components A and B of
\object{HE~1104$-$1805}.  Both the spectra before and after the
correction for telluric features was applied are shown. For clarity,
the uncorrected spectra are shifted down by 0.3 units. The strongest
absorbers are now clear and are marked with the vertical dotted
lines. The atmospheric A and B bands are also marked.}
\label{diff}
\end{figure}

As in W98 and C2000, we also plot the difference between the spectrum
of component A and a scaled version of the spectrum of component B,
that is $f_{\lambda}(A)-c.f_{\lambda}(B)$, where, $c$, is set to
2.8. The difference is plotted in Fig.~\ref{diff}, where we plot it
before and after telluric correction. Apart from a slight excess at
6390\AA\, and the metallic absorption lines in component A, the broad
lines of the quasar disappear and only the excess continuum from
component A remains. The strongest absorbers are now clear and are
marked with the vertical dotted lines.  The wavelengths of the main
absorption lines and the redshifts of the corresponding absorption
line systems are given in Smette et al. (\cite{Smette95}).

\section{Discussion}

The aim of our observations were modest: to measure the redshift of the
lensing galaxy in \object{HE~1104$-$1805}, one of the ever growing list of
lenses with known time delays.  This was possible with only one hour
of VLT time during average weather conditions. The prospects of measuring
the lens redshifts in other systems are therefore very good, at least for
double systems, where aligning the slit simultaneously on the quasar
images and on the lens is straightforward.

With the time delay and the lens and source redshifts known and
several mass models published, it is relatively straight forward to
derive the Hubble constant. W98 present two models to describe the
lens: a singular isothermal sphere with external shear and a singular
isothermal ellipsoid without external shear. For the former (the model
presented by R98 is very similar), one derives $H_0=52$ km/s/Mpc and
for the latter, one derives $H_0=59$ km/s/Mpc. We have set
$(\Omega_M,\Omega_{\Lambda})=(0.3,0.7)$. C2000 also present two models:
a singular ellipsoid without external shear and a two component
model which consists of a singular isothermal ellipsoid to represent
the lensing galaxy and a more extended component representing a galaxy
cluster centered on the lensing galaxy.  The first model leads to
$H_0=62$ km/s/Mpc, but the second leads to values for the Hubble
Constant which are a factor of 1.5 to 2 lower.

\object{HE~1104$-$1805} is somewhat unusual in that the lensing galaxy
is closest to the brighter component. It has been pointed out by
several authors that the lens probably does not consist of a single
galaxy (W98, R98, C2000; Lehar et al. \cite{Lehar00}).  Additional extended mass
is required to match all the observational constraints: the position
of the two images relative to the lens, the magnification ratio of the
images, and the lens orientation and ellipticity. Deep images in the
near-IR may be able to reveal the distribution of this additional
mass.

As the lens appears to be associated with the absorption system seen
in the quasar spectrum at $z=0.728$, one could use the velocity spread
in this system to constrain the lens velocity dispersion. However,
this assumes that the absorption line system and the lens are the same
object or, at the very least, in the same potential well. Without
further supporting evidence that links the absorption line system to
lens, it is premature to use the velocity spread as a constraint on the
mass models.

The lens redshift estimates of W98 and Kochanek et al. (\cite{Koe20})
have proven to be accurate, whereas the estimate of C2000 from IR and
optical photometry is less so. The optical photometry is in good
agreement with the VLT spectrum; however, if the lens is an elliptical
galaxy, then the IR points are over-estimated by 0.3 magnitudes. The
cause of this discrepancy requires further investigation.

\begin{acknowledgements}
  It is a pleasure to thank Thomas Szeifert for his expert support at
  the telescope.  Fr\'ed\'eric Courbin acknowledges financial support
  from the Chilean grant FONDECYT/3990024.  Additional support from
  the European Southern Observatory, through CNRS/CONICYT grant 8730
  ``Mirages gravitationnels avec le VLT: Distribution de mati\`ere
  noire et contraintes cosmologiques''as well as through the TMR
  LENSNET network (ERBFMRXCT97-0172), is also gratefully acknowledged.
\end{acknowledgements}

\end{document}